\documentclass[12pt,preprint]{aastex}

\shorttitle{Morphologies of AGN Host Galaxies}
\shortauthors{Pierce et al.}

\newcommand\q{\par\hangindent 0.5cm\hangafter=1\noindent}

\begin{document}

\title{AEGIS: Host Galaxy Morphologies of X-ray and Infrared--selected AGN at $0.2 \le z < 1.2$}
\author{C.\ M.\ Pierce\altaffilmark{1},
 J.\ M.\ Lotz\altaffilmark{2,3},
 E.\ S.\ Laird\altaffilmark{4},
 L.\ Lin\altaffilmark{5},
 K.\ Nandra\altaffilmark{6},
 J.\ R.\ Primack\altaffilmark{1,7},
 S.\ M.\ Faber\altaffilmark{4},
 P.\ Barmby\altaffilmark{8},
 S.\ Q.\ Park\altaffilmark{8},
 S.\ P.\ Willner\altaffilmark{8},
 S.\ Gwyn\altaffilmark{9},
 D.\ C.\ Koo\altaffilmark{4},
 A.\ L.\ Coil\altaffilmark{10,11},
 M.\ C.\ Cooper\altaffilmark{12},
 A.\ Georgakakis\altaffilmark{6},
 A.\ M.\ Koekemoer\altaffilmark{13},
 K.\ G.\ Noeske\altaffilmark{4},
 B.\ J.\ Weiner\altaffilmark{14},
and C.\ N.\ A.\ Willmer\altaffilmark{10}}

\altaffiltext{1}{Department of Physics, University of California, Santa Cruz, 1156 High Street, Santa Cruz, CA 95064, USA; cpierce@physics.ucsc.edu, joel@scipp.ucsc.edu}
\altaffiltext{2}{National Optical Astronomical Observatories, 950 N.\ Cherry Avenue, Tucson, AZ 85719, USA; lotz@noao.edu}
\altaffiltext{3}{Leo Goldberg Fellow}
\altaffiltext{4}{UCO/Lick Observatory; Department of Astronomy and Astrophysics, University of California, Santa Cruz, 1156 High Street, Santa Cruz, CA 95064, USA; eslaird, faber, koo, kai@ucolick.org}
\altaffiltext{5}{Department of Physics, National Taiwan University, No.\ 1, Sec.\ 4, Roosevelt Road, Taipei 106, Taiwan; lihwai@ucolick.org}
\altaffiltext{6}{Astrophysics Group, Imperial College London, Blackett Laboratory, Prince Consort Rd., London SW7 2AW, UK; k.nandra, a.georgakakis@imperial.ac.uk}
\altaffiltext{7}{Santa Cruz Institute of Particle Physics, University of California, Santa Cruz, 1156 High Street, Santa Cruz, CA 95064, USA}
\altaffiltext{8}{Harvard-Smithsonian Center for Astrophysics, 60 Garden St., Cambridge, MA 02138, USA; pbarmby, spark, willner@cfa.harvard.edu}
\altaffiltext{9}{Department of Physics and Astronomy, University of Victoria, P.O.\ Box 3055, STN CSC, Victoria, BC V8W 3P6, Canada; gwyn@beluga.phys.uvic.ca}
\altaffiltext{10}{Steward Observatory, University of Arizona, 933 N.\ Cherry Avenue, Tucson, AZ 85721, USA; acoil@as.arizona.edu, cnaw@ucolick.org}
\altaffiltext{11}{Hubble Fellow}
\altaffiltext{12}{Department of Astronomy, University of California, Berkeley, Campbell Hall, Berkeley, CA 94720, USA; cooper@astro.berkeley.edu}
\altaffiltext{13}{Space Telescope Science Institute, 3700 San Martin Drive, Baltimore, MD 21218, USA; koekemoe@stsci.edu}
\altaffiltext{14}{Department of Astronomy, University of Maryland, College Park, MD 20742, USA; bjw@ucolick.org}

\begin{abstract}
We visually and quantitatively determine the host galaxy morphologies of 94 intermediate redshift ($0.2 \le z < 1.2$) active galactic nuclei (AGN), selected using {\it Chandra} X-ray and {\it Spitzer} mid-infrared data in the Extended Groth Strip. Using recently developed morphology measures, the second-order moment of the brightest $20\%$ of a galaxy's flux ($M_{20}$) and the Gini coefficient, we find that X-ray--selected AGN mostly reside in E/S0/Sa galaxies ($53^{+11}_{-10}\%$), while IR--selected AGN show no clear preference for host morphology. X-ray--selected AGN hosts are members of close pairs more often than the field population by a factor of $3.3\pm1.4$, but most of these pair members appear to be undisturbed early-type galaxies and do not tend to show direct evidence of gravitational perturbations or interactions. Thus, the activation mechanism for AGN activity remains unknown, even for pair members.
\end{abstract}

\keywords{galaxies: active --- galaxies: interactions --- galaxies: nuclei --- infrared: galaxies --- X-rays: galaxies}

\section{Introduction}
Galaxy interactions and major mergers are currently the most popular explanation for energetically active galactic centers (e.g., Jogee 2005), commonly termed active galactic nuclei (AGN). Both the frequency of disturbed morphologies and the presence of AGN correlate with the host galaxy's IR luminosity (e.g., Sanders \& Mirabel 1996). However, using measures such as concentration, rotational asymmetry ({\it C}, {\it A}; Conselice 2003), and near--neighbor frequency, some studies failed to find a correlation between X-ray--selected AGN and galaxy interactions (e.g., Grogin et al. 2003, 2005, hereafter G03, G05). Thus, galaxy interactions may correlate more strongly with IR--luminous AGN or may be less apparent for X-ray-luminous AGN. Recent galaxy merger simulations also suggest a connection between disturbed morphologies, AGN, and hard X-ray spectra (e.g., DiMatteo, Springel, \& Hernquist 2005, hereafter DSH05; Hopkins et al.\ 2005a, 2005b, hereafter H05a, H05b).

The Gini coefficient, which measures the distribution of flux amongst a galaxy's pixels ({\it G}; Abraham 2003), and the second-order moment of the brightest $20\%$ of a galaxy's flux ($M_{20}$; Lotz, Primack, \& Madau 2004, hereafter LPM04), have been shown to be robust to high $z$ for determining the morphologies of potentially disturbed objects. In particular, $M_{20}$ is more sensitive to merger signatures than other measures of concentration. Morphologically undisturbed galaxies form a well-defined sequence, correlated with Hubble type, in the $G-M_{20}$ plane; merging and interacting galaxies are separated from this sequence by having a high {\it G} for a given $M_{20}$ (LPM04).

In this work, we classify X-ray and IR--selected AGN host galaxy morphologies using {\it G}, $M_{20}$, {\it C}, {\it A}, and visual inspection of {\it Hubble Space Telescope} Advanced Camera for Surveys (ACS) images in the Extended Groth Strip (EGS; Davis et al.\ 2006, hereafter D06; Lotz et al.\ 2006). We also use Deep Extragalactic Evolutionary Probe 2 (DEEP2) spectroscopic velocities to determine the frequency of close kinematic galaxy pairs (Lin et al.\ 2004, 2006). Throughout, we use $H_{0} = 70$ km s$^{-1}$ Mpc$^{-1}$, $\Omega_{\Lambda} = 0.7$, and $\Omega_{M} = 0.3$.

\section{Samples}
We use spectroscopic redshifts from the DEEP2 redshift survey (Davis et al.\ 2003; D06) and photometric redshifts from the Canada--France--Hawaii Telescope Legacy Survey (CFHTLS) Deep Field 3 in {\it ugriz}. At $I < 23.5$, $\sigma_{z,phot} = 0.08(1 + z)$ with a $7\%$ catastrophic failure rate\footnote{See http://astrowww.phys.uvic.ca/grads/gwyn/cfhtls/.}. The inaccuracy of photometric redshifts for very blue objects (i.e., QSOs; Georgakakis et al.\ 2006a) should not significantly affect the present work because such objects are generally excluded due to the compactness of the host galaxies. Photometric redshifts of $I < 23.5$ galaxies agree well with the redshifts implied by IRAC colors (e.g., Barmby et al.\ 2006). The median redshifts of the samples (described below; field galaxies: 0.67, X-ray--selected AGN hosts: 0.74, and IR--selected AGN hosts: 0.46) differ by $\Delta z \le 0.28$.

Our field galaxy sample consists of all EGS galaxies imaged by ACS and meeting the following criteria: $I < 23.5$ (AB), $0.2 \le z < 1.2$, $\langle S/N \rangle$ per pixel $\ge$ 2.5 ({\it G}, $M_{20}$, and {\it C}\/) or $\langle S/N \rangle$ per pixel $\ge$ 4.0 ({\it A}\/), and elliptical Petrosian radius $r_{P} \ge$ 0\farcs3 (Lotz et al.\ 2006). We used the observed {\it V} ({\it I\/}) band images of the field galaxies at $0.2 \le z < 0.6$ ($0.6 \le z < 1.2$) to approximate the rest-frame {\it B} band morphologies. We use {\it G} and $M_{20}$ to classify galaxies in the same manner as the EGS study by Lotz et al.\ (2006). We also use {\it C\/} and {\it A} to identify interacting galaxies ($A\ge0.35$), early-type galaxies ($A < 0.35$, $C \ge 4.0$), and late-type galaxies ($A < 0.35$, $C < 4.0$; Conselice 2003; G03; G05). The second column of Table 1 shows the sizes of the field galaxy, X-ray--selected AGN, and IR--selected AGN samples for various criteria.

Our initial X-ray sample consists of 156 sources from the 200--ks pointing in the Groth--Westphal Strip (Nandra et al.\ 2005) and 409 sources from three 100--ks pointings from early EGS data (Georgakakis et al.\ 2006b) obtained using the {\it Chandra X-ray Observatory} (D06). The on-axis flux limit for hard band--selected sources in the 200--ks (100--ks) observation corresponds to $L_{2-10 \ \rm{keV}} = 2.4\times10^{42}$ erg s$^{-1}$ ($7.2\times10^{42}$ erg s$^{-1}$) at $z = 1$. {\it Chandra} sources were matched to ACS images by requiring a positional offset between the host and the source of $<$ 1\farcs5 for {\it Chandra} off-axis angles (OAAs) $<$5\farcm0 and an offset $<$ 2\farcs0 for OAAs $\ge$5\farcm0. Of the 367 {\it Chandra} sources in the region imaged by the ACS, 80 matched galaxies with $I < 23.5$ at $0.2 \le z < 1.2$, of which $1.5\%$ are expected to be spurious matches. Two of the X-ray sources match a single galaxy, so the sample contains 79 galaxies and 80 AGN. Detection of hard X-ray photons (2-7 keV) and $L_{\rm 2-10 \ keV} > 10^{41}$ erg s$^{-1}$ indicate that all 80 ACS-matched sources are expected to be AGN. Fifty-seven AGN have hosts that meet the S/N and size criteria for $G-M_{20}$ classification. Of these, 37 have high quality spectroscopic redshifts; photometric redshifts are used for the remaining 20. Eleven of the excluded hosts are compact and show no underlying host galaxy, eight have very bright central point sources but are visually classifiable, and four have extended light but low surface brightnesses. (See Table 1.)

Using data from the {\it Spitzer} Infrared Array Camera (IRAC; Fazio et al.\ 2004; D06), we select as AGN the 1,785 sources whose mid-infrared spectral energy distributions follow a power law with a negative spectral index ($f_{\nu}\propto\nu^{\alpha}$; $\alpha < 0$), with a ${\chi}^2$ probability of its being a good fit $P({\chi}^2) > 0.1$ (e.g., Neugebauer et al.\ 1979; Alonso-Herrero et al.\ 2006). These were matched to ACS images by requiring a positional offset of $<$ 2\farcs0. Of the 420 ACS objects matched to IR--selected AGN, 29 have $I < 23.5$ and $0.2 \le z < 1.2$; $2.6\%$ are expected to be spurious matches. Seventeen galaxies meet the S/N and size criteria for $G-M_{20}$ classification. Of these, 7 have high quality spectroscopic redshifts; photometric redshifts are used for the remaining 10. Seven of the excluded hosts are compact and show no underlying host galaxy, four have very bright central point sources but are visually classifiable, and one has low surface brightness. (See Table 1.)

The regions imaged by {\it Chandra}, IRAC, and ACS contain 367 X-ray--selected AGN (199 matched to host galaxies) and 881 IR--selected AGN (372 matched to host galaxies). AGN in 50 of the host galaxies were selected by both X-ray and IR methods. Nine of these 50 AGN hosts have $I < 23.5$ and $0.2 \le z < 1.2$. Of the nine, two match hosts that have $r_{P} \ge 0\farcs3$ and $\langle S/N \rangle$ per pixel $\ge$ 2.5, and seven have bright central point sources, making them too compact for $G-M_{20}$ analysis.

To identify close kinematic galaxy pairs (e.g., Patton et al.\ 2002), we created a spectroscopic sample of galaxies with $I < 23.5$ and spectroscopic redshifts $0.2 \le z < 1.2$. The sample includes 3,139 field galaxies, 48 X-ray--selected AGN hosts, and 11 IR--selected AGN hosts. Using the kinematic pairs identified by Lin et al.\ (2004, 2006) with $\Delta M_{V} \le 2$ ($\Delta M_{I} \le 2$) for $0.2 \le z < 0.6$ ($0.6 \le z < 1.2$), we determined the fractions of AGN hosts and field galaxies with kinematic pairs in our spectroscopic sample. The number of identified pairs is a lower limit as we may miss some companions due to the incompleteness of the spectroscopic sample. The spectroscopic redshift sampling rate among neighbors (separation $<10\farcs0$, $\Delta M_{I} \le 2$, $\Delta M_{V} \le 2$) of AGN hosts is $0.48\pm0.13$, compared to $0.66\pm0.02$ among neighbors of galaxies without AGN. The lower sampling rate around AGN indicates that the differences discussed in \S3 may also be lower limits.

\section{Results}
\subsection{Galaxies Hosting X-ray--selected AGN}
X-ray--selected AGN are mostly found in E/S0/Sa hosts. This is clear from the $G-M_{20}$ classification of the X-ray--selected AGN hosts shown in Figure 1. E/S0/Sa hosts are much more common and Sc/d/Irr hosts are much less common than the field population, Sb/bc hosts are represented at the same rate as the field population, and mergers are more common among AGN hosts by a factor of $2.3\pm0.7$ ($1.7\sigma$ result; Table 1). A two-dimensional K--S test (Fasano \& Franceschini 1987; hereafter FF87) is inconsistent with the {\it G} and $M_{20}$ values of the X-ray--selected AGN hosts and field galaxies being drawn from the same population ($p=8.5\times10^{-8}$). Selecting only galaxies brighter than $M_{B} < -20.5$ (Vega), increases the fraction of E/S0/Sa field galaxies, but the fraction of E/S0/Sa X-ray--selected AGN hosts remains significantly higher (Table 1). Visual classification of the 66 visually classifiable hosts (cf.\ \S2) provides similar results (Table 1). The $C-A$ statistics are consistent with those found by G05. Our finding that most X-ray--selected AGN are hosted by E/S0/Sa galaxies is consistent with previous findings that AGN tend to be hosted by massive ellipticals or bulge-dominated galaxies (e.g., Kauffmann et al.\ 2003; G05). It is also consistent with their red and luminous positions on the color-magnitude diagram (Nandra et al.\ 2006).

X-ray--selected AGN hosted by E/S0/Sa galaxies tend to have softer X-ray spectra\footnote{Hardness ratio: HR $\equiv$ (H-S)/(H+S); H$=2-7$ keV counts; S$=0.5-2$ keV counts. At $z=0$, HR $>-0.25$ ($<-0.25$) indicates high (low) attenuation of soft-band X-rays by gas and dust. Due to redshifting of the energy bands, high-$z$ sources may have harder spectra than observed.} (median HR$=-0.46$) than those hosted by Sc/d/Irr galaxies (median HR$=0.55$), consistent with the expectation that E/S0/Sa galaxies contain less obscuring gas. However, the detection of soft spectra from four X-ray--selected AGN hosted by morphologically disturbed galaxies may indicate surprisingly low levels of obscuration.

Seven of the 48 ($15^{+8}_{-5}\%$) X-ray--selected AGN in the spectroscopic sample are associated with kinematic pairs; in comparison, $4.5\pm0.39\%$ of field galaxies are associated with kinematic pairs. Our results suggest that the X-ray sources have a higher companion rate by a factor $3.3\pm1.4$ ($1.7\sigma$ result). This is greater than that found by G05; however, we find that the companion pairing rate decreases if we only use projected separation to identify pairs. Most X-ray--selected AGN in kinematic pairs are hosted by relatively undisturbed early-type galaxies.

\subsection{Galaxies Hosting IR--selected AGN}
IR--selected AGN inhabit host galaxies of all types, as shown by the $G-M_{20}$ classification of the IR--selected AGN hosts in Figure 2. A two-dimensional K--S test (FF87) is consistent with the IR--selected AGN hosts and field galaxies being drawn from the same population ($p=0.3$). The $C-A$ results similarly indicate little morphological distinction between IR--selected AGN hosts and the field galaxies. Visual classification results suggest a slight shift from Sc/d/Irr hosts to E/S0/Sa hosts (Table 1). Two of the 11 IR--selected AGN in the spectroscopic sample are associated with kinematic pairs, hinting that kinematic pairs may be more common among IR--selected AGN than the field galaxies (cf.\ \S3.1), but this result is statistically weak in our small sample.

\section{Discussion}
The relatively high fraction of X-ray--selected AGN in kinematic pairs supports models that predict that galaxy interactions are responsible for the activation of some galactic nuclei (e.g., DSH05, H05ab). However, most AGN are found in undisturbed, isolated systems, implying that interactions may not be the main method by which AGN are turned on or that the timescales for interactions and AGN activity differ significantly from what is suggested by simulations.

X-ray and IR AGN selection methods select AGN with different host galaxy morphologies. A two-dimensional K--S test (FF87) is inconsistent with the X-ray-- and IR--selected AGN hosts being drawn from the same population ($p=5.4\times10^{-3}$). IR selection via a power-law misses many AGN hosted by E/S0/Sa galaxies; although all X-ray--selected AGN with $I < 23.5$ and $0.2 \le z < 1.2$ are detected by IRAC, most do not have power-law SEDs (see also Barmby et al.\ (2006) who found that only $40\%$ of {\it Chandra} sources have IRAC power laws). X-ray selection misses many AGN hosted by Sc/d/Irr galaxies. Using the AGN with $I < 23.5$ and $0.2 \le z < 1.2$ that are selected by both X-ray and IR methods, $f_{8 \mu{\rm m}}/f_{0.5-7 {\rm keV}} = 2.4\pm0.5$ (consistent with Elvis et al.\ 1994, Table 15); this predicts detectable X-ray emission from the IR--selected AGN, yet many of these objects are not detected by {\it Chandra}. The disparity between the morphologies of X-ray-- and IR--selected AGN suggests that the methods may select AGN at different stages or orientations with varied levels of gas, that galaxy light from the hosts of some X-ray--selected AGN may overwhelm the IR emissions (so that the measured IR emissions do not follow a power-law), or that star-forming field galaxies are contaminating the IR--selected AGN sample.

The observation of soft X-ray spectra from four X-ray--selected AGN hosted by morphologically interacting galaxies may contradict predictions from H05b that soft X-rays from AGN hosted by interacting galaxies should be highly attenuated. However, most of our X-ray--selected AGN are less luminous than the QSO simulated by Hopkins (2006ab). Though it is not surprising that many X-ray--selected AGN hosted by E/S0/Sa galaxies have soft spectra, soft X-ray--selected AGN hosted by interacting galaxies are unexpected. Dissipationless mergers would explain unobscured nuclei but would be unlikely to provide enough gas to trigger AGN. Thus, the observed interactions were probably not responsible for these AGN.

\acknowledgements
This work is based in part on observations made with {\it Hubble Space Telescope}, {\it Spitzer Space Telescope}, which is operated by the Jet Propulsion Laboratory, California Institute of Technology under a contract with NASA, and MegaPrime/MegaCam at the Canada-France-Hawaii Telescope (see D06 for full acknowledgement). We thank the referee for helpful comments. We acknowledge support from NASA grant HST-AR-10675-01-A (CMP); the NOAO Leo Goldberg Fellowship (JML); NSF grants AST-0071198 and AST-0507483, NASA grants HST-GO-10134.18-A, and Chandra-GO5-6141A CalSpace (DCK); and NASA, through Hubble Fellowship grant HF-01182.01-A, awarded by the Space Telescope Science Institute, which is operated by the Association of Universities for Research in Astronomy, Inc., for NASA, under contract NAS 5-26555 (ALC).

\references{}
\q
Abraham, R.\ G., van den Bergh, S., \& Nair, P.\ 2003, ApJ, 588, 218
\q
Alonso-Herrero, A., et al.\ 2006, ApJ, 640, 167
\q
Barmby, P., et al.\ 2006, ApJ, 642, 126
\q
Conselice, C.\ J.\ 2003, ApJS, 147, 1
\q
Davis, M., et al.\ 2003, SPIE, 4834, 161
\q
---.\ 2006, this volume (D06)
\q
Di Matteo, T., Springel, V., \& Hernquist, L.\ 2005, Nature, 433, 604 (DSH05)
\q
Elvis, M., et al.\ 1994, ApJS, 95, 1
\q
Fasano, G., \& Franceschini, A.\ 1987, MNRAS, 225, 155 (FF87)
\q
Fazio, G.\ G., et al.\ 2004, ApJS, 154, 10
\q
Gehrels, N.\ 1986, ApJ, 303, 336
\q
Georgakakis, A., et al.\ 2006a, MNRAS, submitted
\q
---.\ 2006b, this volume
\q
Grogin, N.\ A., et al.\ 2003, ApJ, 595, 685 (G03)
\q
---.\ 2005, ApJ, 627, L97 (G05)
\q
Hopkins, P.\ F., Hernquist, L., Martini, P., Cox, T.\ J., Robertson, B., Di Matteo, T., \& Springel, V.\ 2005a, ApJ, 625, L71 (H05a)
\q
Hopkins, P.\ F., Hernquist, L., Cox, T.\ J., Di Matteo, T., Martini, P., Robertson, B., \& Springel, V.\ 2005b, ApJ, 630, 705 (H05b)
\q
Jogee, S.\ 2005, in AGN Physics on All Scales (Berlin: Springer), in press (astro-ph/0408383)
\q
Kauffmann, G., et al.\ 2003, MNRAS, 346, 1055
\q
Lin, L., et al.\ 2004, ApJ, 617, L9
\q
---.\ 2006, this volume
\q
Lotz, J., et al.\ 2006, ApJ, submitted (astro-ph/0602088)
\q
Lotz, J., Primack, J., \& Madau, P.\ 2004, AJ, 128, 163 (LPM04)
\q
Nandra, K., et al.\ 2005, MNRAS, 356, 568
\q
---.\ 2006, this volume
\q
Neugebauer, G., Oke, J.\ B., Becklin, E.\ E., \& Matthews, K.\ 1979, ApJ, 230, 79
\q
Patton, D.\ R., et al.\ 2002, ApJ, 565, 208
\q
Sanders, D.\ B., \& Mirabel, I.\ F.\ 1996, ARA\&A, 34, 749

\clearpage

\begin{deluxetable}{lcrrrr}
\setlength{\tabcolsep}{0.05in}
\tablewidth{0pc}
\tablecolumns{6}
\tabletypesize{\scriptsize}
\tablecaption{Host Galaxy Morphologies\tablenotemark{a}}
\tablehead{
  \colhead{Sample} & \colhead{\#} & \colhead{Mergers} & \colhead{E/S0/Sa} & \colhead{Sb/bc} & \colhead{Sc/d/Irr} \\
  \colhead{} & \colhead{} & \colhead{($\%$)} & \colhead{($\%$)} & \colhead{($\%$)} & \colhead{($\%$)}
}
\startdata
\cutinhead{$G-M_{20}$ ($I < 23.5$; AB)}
Field & 4435 & 8$\pm$0.4 & 18$\pm$0.6 & 18$\pm$0.6 & 56$\pm$1 \\
X-ray & 57 & 18$^{+7}_{-5}$ & 53$^{+11}_{-10}$ & 16$^{+7}_{-5}$ & 14$^{+7}_{-5}$ \\
IR AGN & 17 & 18$^{+17}_{-10}$ & 18$^{+17}_{-10}$ & 18$^{+17}_{-10}$ & 47$^{+23}_{-16}$ \\
\cutinhead{Visual Classification ($I < 23.5$; AB)}
Field & ... & ... & ... & ... & ... \\
X-ray & 66 & 18$^{+7}_{-5}$ & 47$^{+10}_{-8}$ & 18$^{+7}_{-5}$ & 17$^{+7}_{-5}$ \\
IR AGN & 21 & 24$^{+16}_{-10}$ & 29$^{+17}_{-11}$ & 19$^{+15}_{-9}$ & 29$^{+17}_{-11}$ \\
\cutinhead{$G-M_{20}$ ($M_B < -20.5$; Vega)}
Field & 1504 & 9$\pm$0.8 & 27$\pm$1 & 17$\pm$1 & 47$\pm$2 \\
X-ray & 36 & 14$^{+9}_{-6}$ & 53$^{+15}_{-12}$ & 19$^{+10}_{-7}$ & 14$^{+9}_{-6}$ \\
\cutinhead{$C-A$ ($I < 23.5$; AB)}
Field & 2636 & 3$^{+0.4}_{-0.3}$ & 10$\pm$0.6 & ... & 87$\pm$2 \\
X-ray & 30 & 0$^{+6}_{-0}$ & 30$^{+14}_{-10}$ & ... & 70$^{+19}_{-15}$ \\
IR AGN & 7 & 0$^{+26}_{-0}$ & 0$^{+26}_{-0}$ & ... & 100$^{+54}_{-37}$ \\
\enddata
\tablenotetext{a}{All uncertainties are $1\sigma$ (following Gehrels 1986).}
\end{deluxetable}

\clearpage

\begin{figure}
\plotone{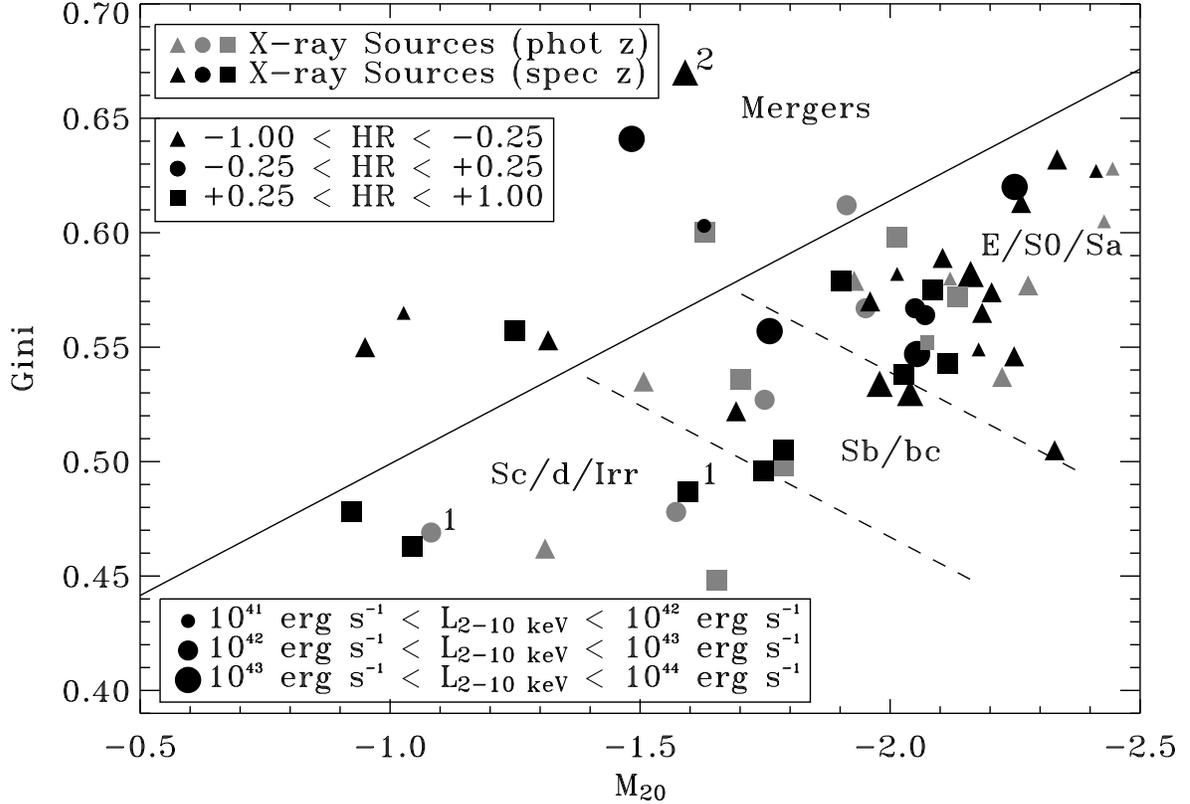}
\caption{$G-M_{20}$ classifications of X-ray--selected AGN hosts. Symbol shape and darkness indicate X-ray hardness ratio and redshift type, respectively. Symbol size indicates $L_{2-10 \ \rm{keV}}$ of the AGN. The symbols marked with `1's are also IR--selected AGN. The symbol marked with a `2' represents a single galaxy associated with two X-ray sources. The solid line roughly separates interacting galaxies from normal galaxies, and the dashed lines separate normal galaxies by Hubble type (Lotz et al.\ 2006). X-ray--selected AGN mostly reside in E/S0/Sa hosts; AGN in Sc/d/Irr hosts tend to have harder spectra than those hosted by E/S0/Sa galaxies.}
\end{figure}

\clearpage

\begin{figure}
\plotone{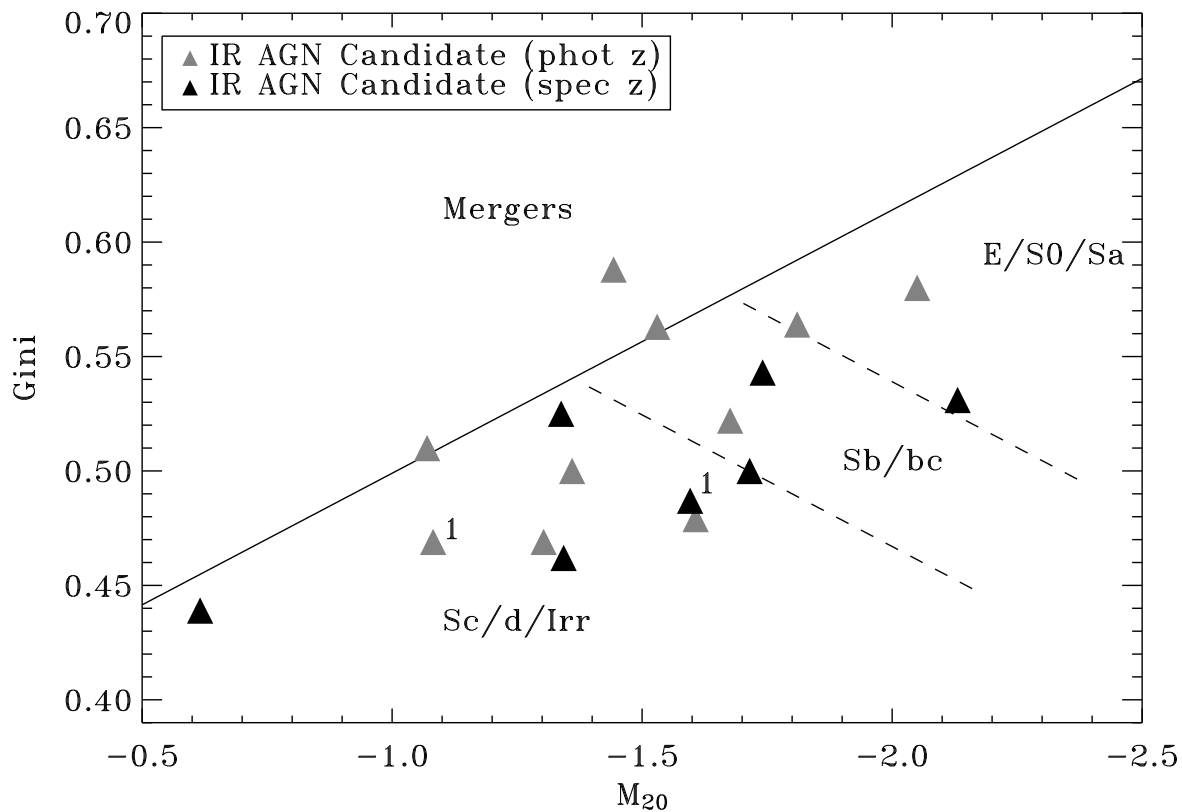}
\caption{$G-M_{20}$ classifications of IR--selected AGN hosts. Symbol darkness indicates redshift type. The symbols marked with `1's are also X-ray--selected AGN. The lines are as in Figure 1. IR--selected AGN show no clear preference for host morphology, though there is a slightly higher fraction of Sc/d/Irr hosts.}
\end{figure}

\end{document}